\documentclass[twocolumn,tighten,times,twocolappendix,numberedappendix,usenames,dvipsnames]{aastex631}

\usepackage{amsmath}
\usepackage{color}
\usepackage{url}
\usepackage{stmaryrd}

\usepackage{etoolbox}

\usepackage{setspace} 
\usepackage{graphicx}
\usepackage[flushleft]{threeparttable}
\usepackage{placeins}
\usepackage{comment}

\usepackage[encapsulated]{CJK}
\usepackage{ucs}
\usepackage[utf8x]{inputenc}

\hypersetup{breaklinks=true,colorlinks=true,linkcolor=blue,citecolor=blue,filecolor=blue,urlcolor=blue}
\makeatletter
\patchcmd{\NAT@citex}
  {\@citea\NAT@hyper@{%
     \NAT@nmfmt{\NAT@nm}%
     \hyper@natlinkbreak{\NAT@aysep\NAT@spacechar}{\@citeb\@extra@b@citeb}%
     \NAT@date}}
  {\@citea\NAT@nmfmt{\NAT@nm}%
   \NAT@aysep\NAT@spacechar\NAT@hyper@{\NAT@date}}{}{}

\patchcmd{\NAT@citex}
  {\@citea\NAT@hyper@{%
     \NAT@nmfmt{\NAT@nm}%
     \hyper@natlinkbreak{\NAT@spacechar\NAT@@open\if*#1*\else#1\NAT@spacechar\fi}%
       {\@citeb\@extra@b@citeb}%
     \NAT@date}}
  {\@citea\NAT@nmfmt{\NAT@nm}%
   \NAT@spacechar\NAT@@open\if*#1*\else#1\NAT@spacechar\fi\NAT@hyper@{\NAT@date}}
  {}{}
  
\usepackage{array}
\newcolumntype{C}[1]{>{\centering\let\newline\\\arraybackslash\hspace{0pt}}m{#1}}

\usepackage{xspace}

\def\aj{AJ}

\def\apj{ApJ}
\def\apjl{ApJ}
\def\apjs{ApJS}
\def\ao{Appl.~Opt.}

\def\aap{A\&A}

\def\baas{BAAS}

\def\mnras{MNRAS}

\def\pasp{PASP}

\def\procspie{Proc.~SPIE}
\def\rmxaa{Rev.~Mex.~Astron.~Astrofis.}

\def\arcdeg{\hbox{$^\circ$}}

\definecolor{burgundy}{rgb}{0.5, 0.0, 0.13}

\usepackage{xspace}

\newcommand{\orcidicon}{\includegraphics[width=0.26cm]{orcid-ID.eps}}

\newcommand{\orcidauthor}[1]{\href{https://orcid.org/#1}{\orcidicon}}

\shorttitle{Spectral variability of the H-poor ejecta in A\,58}
\shortauthors{B.~Montoro-Molina et al.}

\makeatletter
\patchcmd{\frontmatter@RRAP@format}{(}{}{}{}
\patchcmd{\frontmatter@RRAP@format}{)}{}{}{}
\renewcommand\Dated@name{}
\makeatother

\begin{document}

\title{\large Spectral variability of the born-again ejecta in A\,58}

\correspondingauthor{B.\,Montoro-Molina}
\email{borjamm@iaa.es}

\author[0000-0001-9779-4895]{Borja\,Montoro-Molina}
\affil{Instituto de Astrof\'{i}sica de Andaluc\'{i}a, IAA-CSIC, 
Glorieta de la Astronom\'{i}a S/N, E-18008 Granada, Spain}

\author[0000-0002-7759-106X]{Mart\'{i}n \,A.\,Guerrero}
\affil{Instituto de Astrof\'{i}sica de Andaluc\'{i}a, IAA-CSIC, 
Glorieta de la Astronom\'{i}a S/N, E-18008 Granada, Spain}

\author[0000-0002-5406-0813]{Jes\'{u}s\,A.\,Toal\'{a}}
\altaffiliation{Visiting astronomer at the Instituto de Astrof\'{i}sica de Andaluc\'{i}a (IAA-CSIS, Spain) as part of the Centro de Excelencia Severo Ochoa Visiting-Incoming programme.}
\affil{Instituto de Radioastronom\'ia y Astrof\'isica,  UNAM 
Campus Morelia, Apartado postal 3-72, 58090, Morelia, Michoacán, Mexico}

\author[0000-0002-0616-8336]{Janis~B.\,Rodr\'{i}guez-Gonz\'{a}lez}
\affil{Instituto de Radioastronom\'ia y Astrof\'isica,  UNAM 
Campus Morelia, Apartado postal 3-72, 58090, Morelia, Michoacán, Mexico}

\date[]{Submitted to ApJL}

\begin{abstract}
Born-again planetary nebulae (PNe) allow investigating stellar evolution, dust production, and nebular shocks in human timescales. 
Here we present an analysis of multi-epoch optical spectroscopic observations of the born-again PN A\,58 around V605\,Aql, which experienced a {\it very late thermal pulse} (VLTP) about a century ago. 
The H-deficient ejecta has experienced a considerable brightening in the time period considered, from 1996 to 2021, with notable changes also in many emission line ratios. 
Neither the reduction of the extinction caused by the dilution of the ejecta nor the increase of the ionizing photon flux from the central star seem capable to produce these spectral changes, which are instead attributed to shocks in the bipolar H-poor outflow dissociating molecular material and propagating through the outer nebula. 
\end{abstract}


\keywords{\href{https://astrothesaurus.org/uat/847}{Interstellar medium (847)};
\href{https://astrothesaurus.org/uat/1249}{Planetary nebulae (1249)};
\href{http://astrothesaurus.org/uat/1607}{Stellar jets (1607)};
\href{http://astrothesaurus.org/uat/1636}{Stellar winds (1636)};
\href{http://astrothesaurus.org/uat/2050}{Low mass stars(2050)}
\vspace{4pt}
\newline
}

\section{Introduction}
\label{wc:sec:introduction}

Born-again planetary nebulae (PNe) form when the central star of the PN (CSPN) experiences a {\it very late thermal pulse} (VLTP) while descending the white dwarf cooling track \citep[e.g.,][]{Iben1983}. 
Such event injects H-deficient, C-rich material inside the old H-rich PN. 
These objects have proven to be excellent laboratories to study the evolution of low- and intermediate-mass stars, H-deficient environments, shock physics, and dust formation and its reprocessing \citep[e.g.,][and references therein]{Toala2021}.

Born-again PNe have been reported to exhibit great variability. 
Optical and IR studies have demonstrated that their ionization structure and dust properties change in human timescales \citep[][]{Clayton2013,Guerrero2018}. 
For instance, the physical properties of the dust in the youngest born-again PN, the Sakurai's Object, have changed dramatically in only 20 years \citep{Evans2020}.
Meanwhile the ionized H-poor ejecta of the $\sim$1000~yr old born-again PNe A\,30 and A\,78 reveals complex processes of photoevaporation, ablation, and mixing as it interacts with the current fast wind and ionizing photon flux from their CSPNe \citep{Fang2014}.

V605~Aql, the CSPN of A\,58, experienced an outburst in 1919 July 4 \citep{Wolf1920} that was initially classified as a nova event (Nova Aql No.~4).  
Subsequent spectroscopic analyses demonstrated that the progenitor star was instead a H-deficient cool carbon star \citep{Lundmark1921,Bidelman1973}. 
By the mid-90s different authors reported the presence of broad emission of C\,{\sc iv} 5806~\AA\ in V605~Aql, a feature typically found in C-rich [Wolf-Rayet]-type stars with effective temperatures $T_\mathrm{eff}\gtrsim$50,000~K \citep[see, e.g.,][and references therein]{Guerrero1996,Clayton1997,Harrison1996}. 
The H-poor nature of the knot at the center of A\,58 confirmed its born-again nature \citep{Seitter1987,Guerrero1996,Wesson2008}.

\begin{figure*}
\centering
\includegraphics[width=0.5\linewidth]{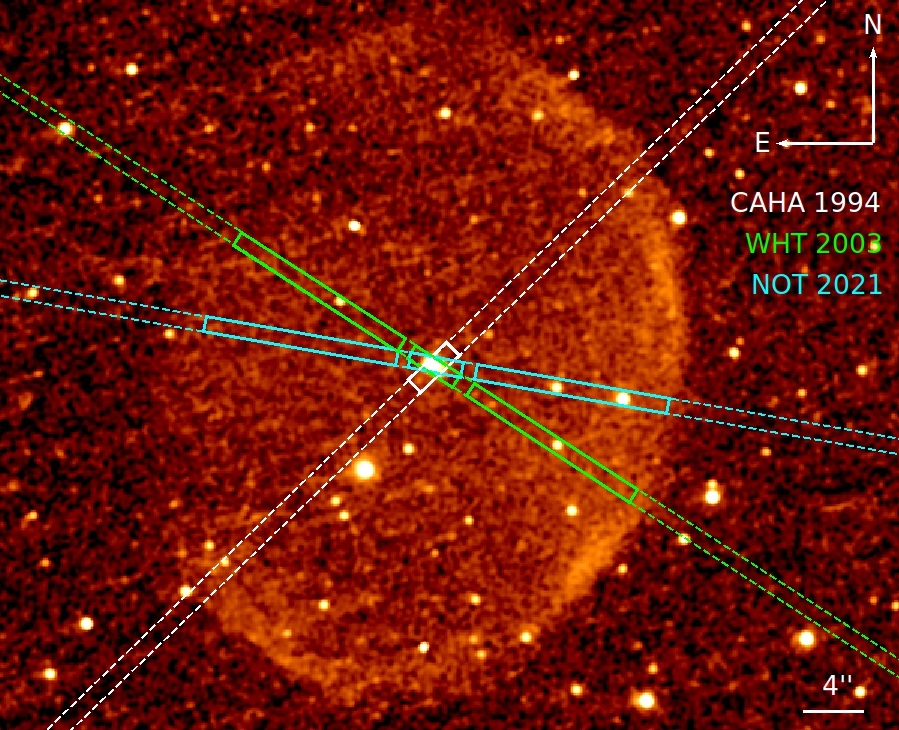}~
\includegraphics[width=0.5\linewidth]{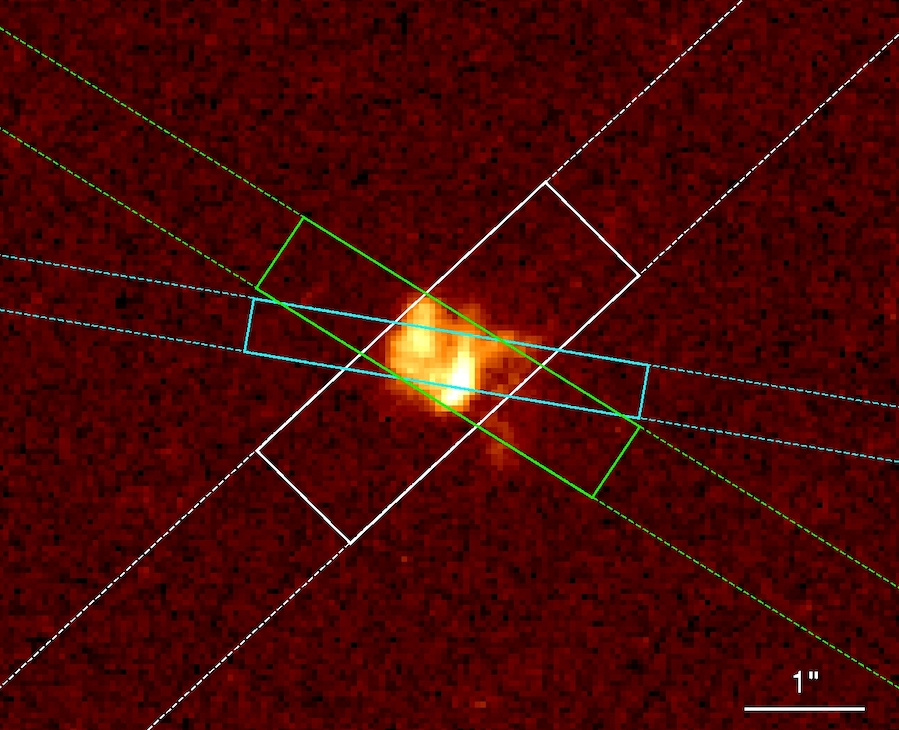}
\caption{
HST WFPC2 WF3 F658N image of A\,58 (left) and HST WFPC2 PC F502N image of its H-deficient central knot (right). 
The dashed lines show the CAHA (white), WHT (green), and NOT (cyan) slits. 
The apertures used for extraction of 1D spectra are overlaid in solid-lines.
}
\label{fig:A58}
\end{figure*}

Time variations of the properties of the H-deficient ejecta around V605~Aql have been  reported. 
Its first detection in radio in 2005 \citep{vanHoof2006} implied a steep emission rise when compared to the previous upper limit reported by \citet{Rao1987}.  
It was attributed either to the optical depth decrease of the ejecta as a consequence of its expansion or to emission increase as V605~Aql is becoming hotter. 
Both possibilities can be true, as multi-epoch narrow-band HST images of A\,58 have shown the expansion of its H-deficient ejecta at a rate of 10 mas~yr$^{-1}$ \citep{Clayton2013}, while detailed non-LTE stellar atmosphere modeling of V605\,Aql have reported an effective temperature of 95,000~K \citep[][]{Clayton2006}.

The H-deficient ejecta of A\,58 has a bipolar morphology, with a disk-like (or torus) structure and a pair of $\sim$1~arcsec in size expanding lobes, more or less with an hourglass shape, oriented along a position angle of $\approx60^{\circ}$ \citep{Clayton2013}. 
\citet{Tafoya2022} recently presented ALMA observations of A\,58 that provide the first image of the molecular content in the ejecta around a born-again star. 
This molecular emission was used to trace the kinematics of the toroidal region and the bipolar outflow in A\,58. 
The former expands radially with a velocity of 100~km~s$^{-1}$, while the bipolar ejection has an expansion velocity of $\approx$280~km~s$^{-1}$ and a dynamical age $\lesssim$20~yr. 
The molecular component is thus younger than the VLTP outburst, suggesting that the bipolar outflow is still been launched.

In this paper we use multi-epoch optical spectroscopic observations of the old H-rich nebula A\,58 and its recent H-deficient ejecta to investigate spectral time variations.
The observations are described in Section~2, the results on the spectral evolution and its possible origins are presented in Sections~3 and 4, respectively, and a final remark is presented in Section~5.

\section{Observations and data preparation}
\label{sec:observations}

We have gathered optical long-slit intermediate-dispersion spectroscopic observations of A\,58 from three different epochs spanning from 1994 to 2021. 
These include two data sets of previously published observations obtained with the Twin Cassegrain Spectrograph (TCS) at the 3.5 m telescope of the Calar Alto Astronomical Observatory (CAHA) on 1994 July 26--28 (1994.57) and with the Intermediate-dispersion Spectrograph and Imaging System (ISIS) at the 4.2 m William Herschel Telescope (WHT) of the Observatorio de El Roque de los Muchachos (ORM) on 2003 August 1 (2003.58). 
Details of these observations were presented by \citet{Guerrero1996} and \citet{Wesson2008}, respectively\footnote{
We note that the CAHA observations are proprietary, whereas the WHT observations are public and were downloaded from the Isaac Newton Group Archive at \url{http://casu.ast.cam.ac.uk/casuadc/ingarch}.}. 
The slit position angles (PAs), widths and spectral resolutions of the CAHA observations were 134$^{\circ}$, 1.2 arcsec, and 4 \AA\ and 3.5 \AA\ for the TCS blue and red arms, respectively, and those of the WHT observations were 57$^{\circ}$, 0.78 arcsec, and 2.3 \AA\ and 3.5 \AA\ for the ISIS blue and red arms, respectively. 

We observed A\,58 on 2021 June 11 (2021.44) with the ALhambra Faint Object Spectrograph and Camera (ALFOSC) mounted on the 2.5 m Nordic Optical Telescope (NOT) of the ORM. 
The E2V CCD detector was used, providing a spatial scale of 0.21 arcsec pix$^{-1}$.
Three 1800~s exposures were obtained with Grism \#7, which has a dispersion of 1.7~\AA~pix$^{-1}$ and a spectral range 3650--7110 \AA. 
The slit was placed at a PA of 80$^{\circ}$ and its width was set at 0.5~arcsec, providing a spectral resolution of 4.2 \AA. 
The average seeing of the night was $\sim$1~arcsec. 
All observations were processed following IRAF standard routines \citep{Tody1986, Tody1993}.

The slit positions of the  observations are presented in Figure~\ref{fig:A58}, where the left panel presents an HST [N~{\sc ii}] narrow-band image of A\,58 obtained on 2001 May 27 (2001.40, Prop.\,ID.: 90920; PI: K.\,Hinkle), whereas the right panel presents a zoom-in view of the central knot from an HST [O~{\sc iii}] narrow-band image obtained on 2009 March 19 (2009.21, Prop.\,ID.: 11985; PI: G.\,Clayton).

\section{Spectral time evolution of A\,58}
\label{sec:results}

The comparison of multi-epoch spectroscopic observations requires a careful cross-calibration and assessment of the observation conditions (sky transparency, seeing, ...) and instrumental configurations (slit width, slit position angle, spectral resolution, spatial scale, signal-to-noise ratio, ...)
To gauge all these effects in the analysis of the spectral time evolution of A\,58, we will investigate first the outer nebula, where the intensity of emission lines is expected to vary only on long timescales \citep{Guerrero2018}, and then proceed to the analysis of its central knot.

\begin{figure*}
\centering
\includegraphics[width=1\linewidth]{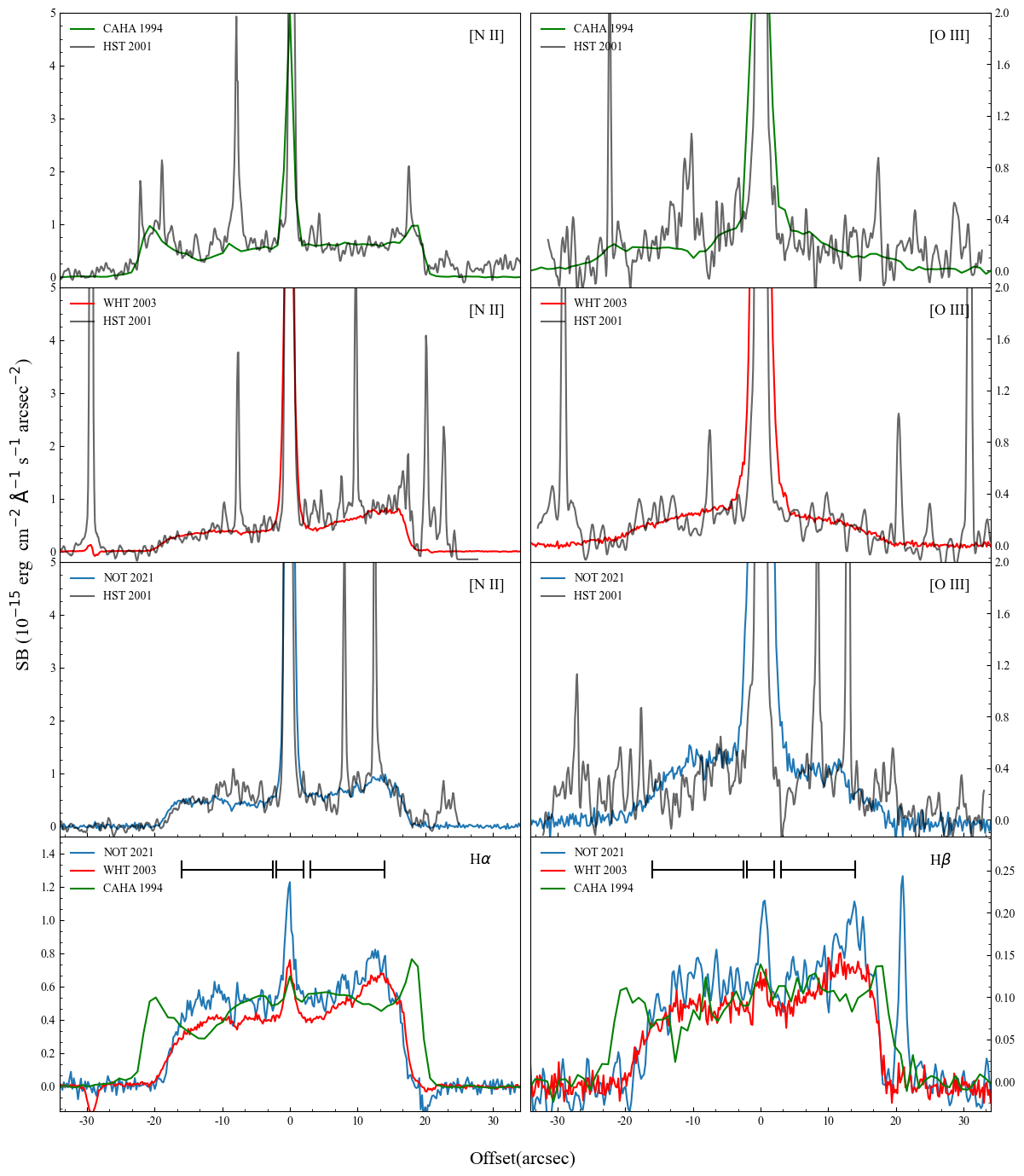}
\caption{
Comparison of SB profiles of the HST WFPC2 F502N and F658N (2001.44), and CAHA (1994.57), WHT (2003.58), and NOT (2021.44) H$\beta$, H$\alpha$, [O~{\sc iii}] $\lambda$5007 and [N~{\sc ii}] $\lambda$6584. 
The CAHA H$\alpha$ and [N~{\sc ii}] SB profiles (green in the top-left and bottom-left panels) extracted from the red arm have been scaled by 0.65 to match the HST WFPC2 F658N SB profile. 
The apertures used to extract the spectra are shown in solid black lines.
}
\label{fig:A58_profiles}
\end{figure*}

These analyses will make use of H$\beta$, H$\alpha$, [O~{\sc iii}] $\lambda$5007 and [N~{\sc ii}] $\lambda$6584 surface brightness (SB) profiles extracted from the 
spectroscopic observations.  
SB profiles along the slits of the spectroscopic observations have also been extracted from the HST WFPC2 WF3 [O~{\sc iii}] and [N~{\sc ii}] narrow-band images for comparison. 
The SB profiles are shown in Figure~\ref{fig:A58_profiles}, where we note the low S/N ratio of the HST [O~{\sc iii}] SB profile.

\begin{table*}
\footnotesize
\begin{center}
\caption{De-reddened line fluxes relative to $F$(H$\alpha)$=1 and physical parameters obtained for the three spectroscopic observations of A\,58. 
}
\label{tab:ratios_total}
\begin{tabular}{lccccc|cc}
\hline
          & \multicolumn{5}{c}{Central knot}  & \multicolumn{2}{c}{Outer nebula} \\
\hline
          & \multicolumn{1}{c}{CAHA}  &  \multicolumn{2}{c}{WHT}    & \multicolumn{2}{c}{NOT} 
          &  WHT  & NOT \\
          & \multicolumn{1}{c}{TCS}   &  \multicolumn{2}{c}{ISIS}   & \multicolumn{2}{c}{ALFOSC}  &  ISIS  & ALFOSC \\
          & \multicolumn{1}{c}{(July 1994)}&  \multicolumn{2}{c}{(August 2003)} & \multicolumn{2}{c}{(June 2021)}  &  (August 2003) & (June 2021) \\
\hline
c(H$\beta$)  &  1.15 & 1.21$\pm$0.20 & 1.15 & 1.07$\pm$0.19 & 1.15  & 0.57$\pm$0.07 & 0.64$\pm$0.06\\  
\hline
{[O~{\sc ii}] } 3727 	&	$\dots$~~	&	38$\pm$7	&	34$\pm$7	&	104$\pm$20	&	115$\pm$23	&	77$\pm$2	&	$\dots$~~	\\
{{[Ne~{\sc iii}]} } 3869 	&	$\dots$~~	&	29$\pm$5	&	27$\pm$5	&	155$\pm$27	&	170$\pm$32	&	12.5$\pm$0.8	&	$\dots$~~	\\
{H$\zeta$ + He~{\sc i} } 3889 	&	$\dots$~~	&	$\dots$~~	&	$\dots$~~	&	2.0$\pm$0.5	&	2.2$\pm$0.6	&	4.6$\pm$0.7	&	$\dots$~~	\\
{H$\epsilon$ + [Ne~{\sc iii}]  } 3969   	&	$\dots$~~	&	12$\pm$2	&	11$\pm$2	&	50$\pm$9	&	54$\pm$10	&	7.3$\pm$0.7	&	$\dots$~~	\\
{{[S~{\sc ii}]} } 4069    	&	$\dots$~~	&	0.8$\pm$0.2	&	0.7$\pm$0.2	&	1.1$\pm$0.3	&	1.2$\pm$0.3	&	$\dots$~~	&	$\dots$~~	\\
{H$\delta$} 4101 	&	$\dots$~~	&	$\dots$~~	&	$\dots$~~	&	$\dots$~~	&	$\dots$~~	&	7.3$\pm$0.5	&	$\dots$~~	\\
{H$\gamma$  } 4340  	&	$\dots$~~	&	$\dots$~~	&	$\dots$~~	&	$\dots$~~	&	$\dots$~~	&	16.1$\pm$0.6	&	15.6$\pm$0.4	\\
{{[O~\sc{iii}]}   } 4363  	&	$\dots$~~	&	7.3$\pm$1.2	&	6.9$\pm$1.1	&	16$\pm$2	&	17$\pm$3	&	$<$0.3	&	$<$0.09	\\
{He~\sc{i}      } 4471     	&	$\dots$~~	&	0.5$\pm$0.1	&	0.5$\pm$0.1	&	0.9$\pm$0.2	&	0.9$\pm$0.2	&	$\dots$~~	&	$\dots$~~	\\
{He~\sc{ii} } 4686	&	$\dots$~~	&	0.4$\pm$0.1	&	0.3$\pm$0.1	&	1.3$\pm$0.2	&	1.4$\pm$0.2	&	9.3$\pm$0.4	&	6.4$\pm$0.2	\\
{{[Ar~{\sc iv}]}} 4711	&	$\dots$~~	&	0.4$\pm$0.1	&	0.4$\pm$0.1	&	1.1$\pm$0.2	&	1.3$\pm$0.2	&	$\dots$~~	&	$\dots$~~	\\
{{[Ar~{\sc iv}]}} 4740    	&	$\dots$~~	&	0.7$\pm$0.1	&	0.70$\pm$0.1	&	1.0$\pm$0.2	&	1.1$\pm$0.2	&	$\dots$~~	&	$\dots$~~	\\
{H$\beta$   } 4861   	&	$\dots$~~  &	0.4$\pm$0.1	&	0.4$\pm$0.1	&	0.4$\pm$0.1	&	0.4$\pm$0.1	&	32.9$\pm$0.8	&	32.9$\pm$0.5	\\
{{[O~\sc{iii}]}     } 4959    	&	27$\pm$5	&	67$\pm$10	&	64$\pm$10	&	172$\pm$25	&	181$\pm$26	&	31.4$\pm$0.7	&	31.7$\pm$0.4	\\
{{[O~\sc{iii}]}    } 5007    	&	85$\pm$15   &	205$\pm$31	&	196$\pm$30	&	512$\pm$74	&	537$\pm$78	&	98$\pm$2	&	108$\pm$2	\\
{{[N~{\sc i}]}   } 5199   	&	$\dots$~~	&	1.7$\pm$0.3	&	1.6$\pm$0.3	&	1.2$\pm$0.2	&	1.2$\pm$0.2	&	$\dots$~~	&	$\dots$~~	\\
{{[N~{\sc ii}]} } 5755	&	0.6$\pm$0.1	&	0.8$\pm$0.2	&	0.8$\pm$0.2	&	0.9$\pm$0.2	&	0.9$\pm$0.2	&	$<$0.07	&	$<$0.03	\\
{C~{\sc iv}      } 5806      	&	8.1$\pm$1.6	&	11$\pm$2	&	11$\pm$2	&	10$\pm$2	&	10$\pm$2	&	$\dots$~~	&	$\dots$~~	\\
{He~{\sc i}     } 5876      	&	0.8$\pm$0.2	&	2.0$\pm$0.4	&	1.9$\pm$0.4	&	2.7$\pm$0.5	&	2.8$\pm$0.5	&	6.2$\pm$0.2	&	$\dots$~~	\\
{{[O~{\sc i}]}      } 6300     	&	8.9$\pm$2.0	&	11$\pm$2	&	11$\pm$2	&	13$\pm$3	&	13$\pm$3	&	$\dots$~~	&	$\dots$~~	\\
{{[O~{\sc i}]}     } 6363    	&	2.9$\pm$0.7	&	4.0$\pm$0.8	&	4.0$\pm$0.8	&	4.8$\pm$0.9	&	4.8$\pm$1.0	&	1.2$\pm$0.1	&	$\dots$~~	\\
{{[N~\sc{ii}]}    } 6548   	&	6.0$\pm$1.4	&	8.6$\pm$1.8	&	8.7$\pm$1.9	&	10$\pm$2	&	10$\pm$2	&	34$\pm$1	&	28.6$\pm$0.5	\\
{H$\alpha$        } 6563       	&	1.0$\pm$0.2	&	1.0$\pm$0.2	&	1.0$\pm$0.2	&	1.0$\pm$0.2	&	1.0$\pm$0.2	&	100$\pm$3	&	100$\pm$2	\\
{{[N~\sc{ii}]}     } 6584     	&	19$\pm$5    &	25$\pm$5	&	25$\pm$6	&	31$\pm$6	&	31$\pm$7	&	100$\pm$3	&	90$\pm$2	\\
{He~{\sc i}     } 6678     	&	0.4$\pm$0.1	&	0.6$\pm$0.1	&	0.6$\pm$0.1	&	0.7$\pm$0.1	&	0.7$\pm$0.1	&	1.3$\pm$0.1	&	$\dots$~~	\\
{{[S~\sc{ii}]}    } 6717   	&	0.5$\pm$0.1	&	1.0$\pm$0.2	&	1.0$\pm$0.2	&	1.0$\pm$0.2	&	1.0$\pm$0.2	&	18.2$\pm$0.9	&	14.4$\pm$0.3	\\
{{[S~\sc{ii}]}    } 6731    &	0.9$\pm$0.2	&	1.4$\pm$0.3	&	1.4$\pm$0.3	&	1.4$\pm$0.3	&	1.4$\pm$0.3	&	12.9$\pm$0.4	&	10.4$\pm$0.2	\\
{He~{\sc i}      } 7065     	&	1.1$\pm$0.3	&	1.0$\pm$0.2	&	1.0$\pm$0.3	&	1.0$\pm$0.2	&	1.0$\pm$0.2	&	$\dots$~~	&	$\dots$~~	\\
{C~{\sc ii} 7064 +C~{\sc i} } 7066    	&	1.0$\pm$0.3	&	1.0$\pm$0.2	&	1.0$\pm$0.2	&	1.0$\pm$0.2	&	1.0$\pm$0.2	&	$\dots$~~	&	$\dots$~~	\\
{{[Ar~\sc{iii}]}    } 7135      	&	2.2$\pm$0.6	&	1.7$\pm$0.4	&	1.7$\pm$0.4	&	$\dots$~~	&	$\dots$~~	&	3.7$\pm$0.1	&	$\dots$~~	\\
{{[O~\sc{ii}]}    } 7320   	&	10$\pm$3	&	6.3$\pm$1.6	&	6.4$\pm$1.6	&	$\dots$~~	&	$\dots$~~	&	$\dots$~~	&	$\dots$~~	\\
{{[O~\sc{ii}]}    } 7330    	&	8.6$\pm$2.3	&	5.3$\pm$1.3	&	5.4$\pm$1.4	&	$\dots$~~	&	$\dots$~~	&	$\dots$~~	&	$\dots$~~	\\
{C~{\sc iv}    } 7724     	&	$\dots$~~	&	1.1$\pm$0.3	&	1.1$\pm$0.3	&	$\dots$~~	&	$\dots$~~	&	$\dots$~~	&	$\dots$~~	\\

\hline
log$_{10}(F$(H$\alpha$)$)$  & $-$15.3 & \multicolumn{2}{c}{$-$15.0} &  \multicolumn{2}{c}{$-$14.80} & $-$13.97 & $-$13.88 \\
SB ($10^{-16}$ erg~cm$^{-2}$~s$^{-1}$~arcsec$^{-1}$) & 0.6 & $\dots$ & 3.9  & $\dots$ & 8.0 & 6.7  & 8.9 \\
\hline

$T_\mathrm{e}$([N {\sc ii}])       &  14200$\pm$100  &  14700$\pm$900  & 14400$\pm$900  &  13700$\pm$600  &  13800$\pm$600  & $<$4400 &  $<$3900 \\
$T_\mathrm{e}$([O {\sc iii}])      & $\dots$ &  21200$\pm$300 & 20800$\pm$300  &  19200$\pm$200  &  19200$\pm$200  & $<$8100 & $<$6100 \\ 
$n_\mathrm{e}$([{S {\sc ii}}])     &  2200$\pm$2000 &  2500$\pm$200  &   2500$\pm$200   &  2700$\pm$100 & 2800$\pm$100 &  $<$100 &  $<$100 \\ 
$n_\mathrm{e}$([{Ar {\sc iv}}])    & $\dots$ & $\dots$ & $\dots$ & 2600$\pm$200 &   2400$\pm$200 & $\dots$ & $\dots$ \\
\hline
\end{tabular}
\end{center}
\end{table*}

\subsection{The outer nebula}
\label{sec:outer_neb}

The comparison of the HST, WHT and NOT [N~{\sc ii}] and [O~{\sc iii}] SB profiles in Figure~\ref{fig:A58_profiles} reveals an excellent match, confirming that there are no drastic SB variations of the outer shell of A\,58 in the 2001.40 to 2021.44 period covered by these observations.  
This is also the case for the HST and CAHA [O~{\sc iii}] SB profiles, but not for the [N~{\sc ii}] ones, with a larger SB for the CAHA data. 
A similar SB mismatch is found for other emission lines derived from the CAHA TCS red arm spectrum with respect to the WHT and NOT spectra in nebular regions around the central knot where the SB levels were expected to be similar.  
We therefore decided to scale down the flux of the CAHA TCS red arm data, which will be not used to investigate SB variations.

The comparison of the CAHA, WHT and NOT H$\beta$, H$\alpha$, [O~{\sc iii}] and [N~{\sc ii}] SB profiles in Figure~\ref{fig:A58_profiles} indicates that the WHT and NOT slits register regions of the outer nebula of similar SB and excitation, but not the CAHA one, with notably different spatial extent and SB variations. 
The latter would then be excluded for the investigation of the possible time evolution of emission line ratios in the outer nebula.

Spectra of the outer nebula were extracted from the WHT and NOT apertures shown in Figure~\ref{fig:A58} and in the bottom panels of Figure~\ref{fig:A58_profiles}, which have similar SB profiles in different emission lines. 
The line fluxes were then measured using the {\sc IRAF} routine \emph{splot} and their errors computed following the scheme described by \citet{Tresse1999}.

The H$\beta$ and H$\alpha$ fluxes were corrected from the contribution of the He~{\sc ii} $\lambda\lambda$4860,6560 Pickering lines scaling their intensities to that of He~{\sc ii} $\lambda$4686 by 0.052 and 0.136, respectively \citep{HS87}.  
These were then used to derive the value of the logarithmic extinction coefficient, $c$(H$\beta$), adopting a recombination Case B \citep{Osterbrock2006}. 
The theoretical value of the H$\alpha$/H$\beta$ ratio depends on the electron temperature ($T_{\rm e}$), but the $T_{\rm e}$ sensitive [O~{\sc iii}] $\lambda$4363 and [N~{\sc ii}] $\lambda$5755 auroral lines are not detected 
and only upper limits can be set.  
Using an iterative method for $T_{\rm e}$ and $c$(H$\beta$), we estimated upper limits for $T_{\rm e}$ [N~{\sc ii}] $<$3,900~K and $T_{\rm e}$ [O~{\sc iii}] $<$6,100~K. 
An intermediate value of 5,000~K was thus assumed for a theoretical value of 3.04 for H$\alpha$/H$\beta$.  
We note that the values of $T_{\rm e}$ are notably low, but this seems to be the trend for the outer nebulae of born-again PNe as a consequence of the reduction in the stellar UV flux after the VLTP event \citep{Guerrero2018}.

The values of $c$(H$\beta$) are then estimated to be 0.57$\pm$0.07 and 0.64$\pm$0.06 for the WHT and NOT data, respectively, which can be considered to be consistent among then. 
Assuming that the extinction is purely interstellar, it can be used to assess the distance to A\,58.  
According to Bayestar19\footnote{\url{http://argonaut.skymaps.info}} \citep{Green2018}, the extinction along the direction of A\,58 varies from $E(g-r$)=0.31$\pm$0.02 mag at 0.88 kpc to 0.39$\pm$0.03 mag at 3.89 kpc, and then increases to 0.47$\pm$0.02 mag for distances above 6.20 kpc.  
The conversion from $E(g-r$) to $E(B-V$) \citep{SF11} implies $c$(H$\beta$) in the range 0.45 to 0.57 for distances in the range 0.88 to 3.89 kpc, and 0.69 for distances above 6.20 kpc, which bracket the values reported here, but dismiss the lower value of 0.29 and higher value of 1.04 reported previously by \citet{Guerrero1996} and \citet{Wesson2008}, respectively. 
The reddening-distance diagram derived from Bayestar19 and the extinction derived from the WHT and NOT data imply a distance of 4.3$_{-0.6}^{+1.7}$ kpc towards A\,58.

The dereddened line intensities of the outer nebula of A\,58 derived from the WHT ISIS and NOT ALFOSC spectra are listed in the rightmost columns of  Table~\ref{tab:ratios_total}. 

\subsection{The central knot}

Spectra of the central knot of A\,58 were extracted from the CAHA, WHT, and NOT data using the apertures shown in Figure~\ref{fig:A58} and in the bottom panels of Figure~\ref{fig:A58_profiles}.
Note that the faint emission of the H~{\sc i} Balmer and He~{\sc i} and He~{\sc ii} lines from the VLTP ejecta is diluted by the bright emission from the H-rich old nebula \citep[see, for instance, the sophisticated analysis required by the emission of the born-again ejecta of HuBi\,1 presented by][]{Borja2022}. 
This contamination affects specially the CAHA spectrum that used a wider long-slit. 
To remove the contribution of the emission from the outer H-rich nebula, an average spectrum of the outer nebula was extracted from apertures adjacent to that of the central knot and subtracted from the central knot spectrum after scaling it according to their relative aperture sizes. 

The central knot line fluxes were measured from its background-subtracted spectrum as described above for the outer nebula. 
The fluxes of the H$\alpha$ and H$\beta$ emission lines were also computed from the SB profiles presented in Figure~\ref{fig:A58_profiles} and found consistent with those measured from the central knot spectra. 
As described in the previous section, the H$\beta$ and H$\alpha$ line fluxes were corrected from the contribution of the He~{\sc ii} $\lambda\lambda$4860,6560 Pickering lines.

The intensities of the H$\alpha$ and H$\beta$ emission lines were then used to derive the value of $c$(H$\beta$) adopting also a recombination Case B and following the iterative method with $T_{\rm e}$ described above. 
In this case, a theoretical $I$(H$\alpha$)/$I$(H$\beta$) flux ratio of 2.76 was adopted, as appropriate for a $T_{\rm e}$ value of 20,000~K (see Table \ref{tab:ratios_total}). 
Values of $c$(H$\beta$) for the knot of A\,58 are found to be 
1.21$\pm$0.20 and 1.07$\pm$0.19 for the WHT and NOT observations, respectively. 
The decrease of the value of $c$(H$\beta$) in the time period from 2003.58 to 2021.44 is not considered to be significant given the uncertainties of each individual value.
The intrinsic line intensities of the central knot of A\,58 referred to a value of unity for H$\alpha$ are presented in Table~\ref{tab:ratios_total}. 
The line intensities of the WHT and NOT spectra were dereddened using the value of $c$(H$\beta$) derived for each one, but also using an average value of 1.15. 
The latter is used for the CAHA TCS spectrum due to its red arm calibration uncertainty.

\begin{figure}
\centering
\includegraphics[width=\linewidth]{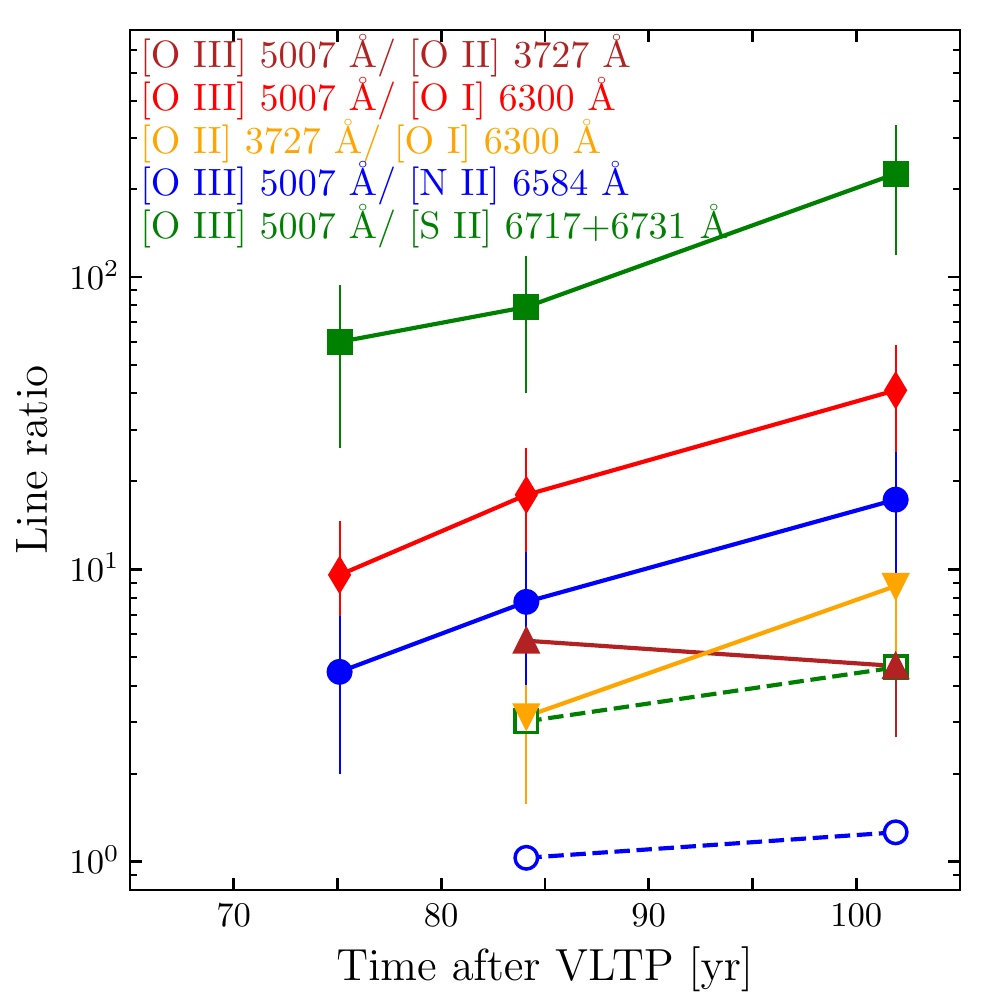}
\caption{
Variation with time of selected line ratios of A\,58 for its H-deficient knot (filled symbols, solid lines) and H-rich outer nebula (open symbols, dashed lines).  
Symbol shapes and colors represent different ratios. 
Time is measured from the VLTP event of V605~Aql in July 1919.
}
\label{fig:histo}
\end{figure}


\subsection{Spectroscopic evolution} 

The emission line intensities and H$\alpha$ fluxes and SBs of the outer H-rich nebula and the H-deficient knot of A\,58 derived from multi-epoch optical spectroscopic observations spanning from 1994 and 2021, including CAHA 1994.57, WHT 2003.58, and NOT 2021.44 data, are presented in Table~\ref{tab:ratios_total}.  
These correspond to 75.1, 84.1, and 101.9 yrs after the VLTP event, respectively. 
Information on the [O~{\sc iii}] $\lambda$5007 and [N~{\sc ii}] $\lambda$6584 emission lines is also provided by the 2001.40 (81.9 yrs) and 2009.21 (89.7 yrs) HST WFPC2 data. 
Different line intensity ratios derived from these data sets are presented in Figure~\ref{fig:histo}, that will be used in conjunction with the SB profiles shown in Figure~\ref{fig:A58_profiles} to investigate the spectral variability of the outer H-rich nebula of A\,58 and the H-deficient ejecta near its central star V605~Aql. 

The $\approx$30\% increase of the H$\alpha$ SB of the outer shell of A\,58 in the time period 
between 2003.58 and 2021.44 shown in Table~\ref{tab:ratios_total} and Figure~\ref{fig:A58_profiles} is suggestive of an apparent overall nebular brightening. 
We note, however, that it might be within the cross-calibration uncertainty and certainly affected by the different nebular regions probed in these two epochs.  
On the other hand, the line intensities in Table~\ref{tab:ratios_total} indicate moderate $\approx$10\% increase of [O~{\sc iii}] and decline of [N~{\sc ii}] in the same time period, with larger $\approx$25\% reductions of the He~{\sc ii} and [S~{\sc ii}] emission lines. 
These trends are illustrated for the [O~{\sc iii}]/[N~{\sc ii}] and [O~{\sc iii}]/[S~{\sc ii}] line ratios in Figure~\ref{fig:histo}.  

On the other hand, the emission line intensities, the H$\alpha$ SB, and the SB profiles of the H-deficient knot reveal unquestionable changes with time.
Its H$\alpha$ emission has brightened by a $\approx$13 factor, whereas the line intensity ratios imply also significant variations, with a general increase of all line intensity ratios with respect to H$\alpha$. 
This emission enhancement affects notably the high-excitation emission lines of [Ne~{\sc iii}], He~{\sc ii}, and [O~{\sc iii}].  
As for the low-excitation emission lines, the [O~{\sc ii}] emission lines increases its intensity notably from 2003 to 2021, whereas those of [N~{\sc i}] and [S~{\sc ii}] $\lambda\lambda$6716,6731 present marginal 5\%--10\% decrements. 
These large variations reflect in the obvious trends shown in Figure~\ref{fig:histo}.  

\subsection{Comparison with previous spectroscopic information}

As noted in Section~3.1, the values of $c$(H$\beta$) derived for the outer shell here and those reported by \citet{Guerrero1996} and \citet{Wesson2008} differ notably.  
This is also the case for the central knot, for which our value of $c$(H$\beta$) of 1.15 is notable smaller than that of 2.0 reported by \citet{Wesson2008}. 

A detailed comparison with the line ratios presented by these authors confirms the issue with the flux calibration of the CAHA TCS red arm of the data set used by \citet{Guerrero1996} reported in Section~3.1.  
The line ratios of the outer shell among the different works may differ up to 25\%, which can be considered otherwise generally consistent given the different apertures used to extract the spectra here and in these works.  
The latter included the low-excitation nebular edge that was purposely excluded here to allow a multi-epoch comparison.
As for the central knot, the dependence with the definition of the spectral aperture, which is not provided by these authors, is even more critical, given its small angular size.  
Moreover the determination of the H$\beta$ and H$\alpha$ line fluxes relies strongly on the capability to remove the contribution of their bright emission from the outer shell to that of the central knot.

A careful double-check of the flux calibration used here for the WHT ISIS and NOT ALFOSC data against the spectro-photometric standard stars has been successfully 
passed.  
Therefore, it can be concluded that, besides the issue in the CAHA TCS red arm calibration discovered here, the difference in the line intensity ratios fluxes presented in this paper with respect to those presented by \citet{Guerrero1996} and \citet{Wesson2008} can be interpreted in terms of the different spectral apertures defined by each author and the method used to subtract the nebular emission of the outer shell from the central knot. 

\section{Origins of the variations}

The analysis presented in the previous section implies that the emission from the outer shell of A\,58 presents only subtle variations if any in the time period investigated, whereas its central H-deficient knot has brightened dramatically and experienced notable spectral variations.  
Its emission has indeed brightened in the period 1994.57 to 2021.44 by factors $\approx$13 for H$\alpha$ and $\gtrsim$80 for [O~{\sc iii}], in line with the radio brightening observed since 2005 \citep{vanHoof2006}.

Three different non-exclusive mechanisms can be argued to produce these variations: 
{\it i}) the expansion of the H-deficient ejecta can be expected to reduce its optical thickness and, as it becomes more transparent with time, the extinction will also reduce, 
{\it ii}) the effective temperature increase of the CSPN of A\,58 results in a much larger ionizing photon flux that changes the ionization balance, and 
{\it iii}) the shocks associated with the expansion of the H-deficient ejecta reach a larger volume of gas with time. 

\subsection{Dilution of the Ejecta} 

The values of the logarithmic extinction $c$(H$\beta$) obtained for the WHT (2003.58) and NOT (2021.44) observations are consistent within their error values.  
It seems that a reduction of the extinction can not be invoked to fully explain the observed SB variations.  
Furthermore, the variations in ionization stage would be difficult to explain by changes in the transparency of the ejecta, neither the radio brightening.

\begin{figure}
\centering
\includegraphics[width=\linewidth]{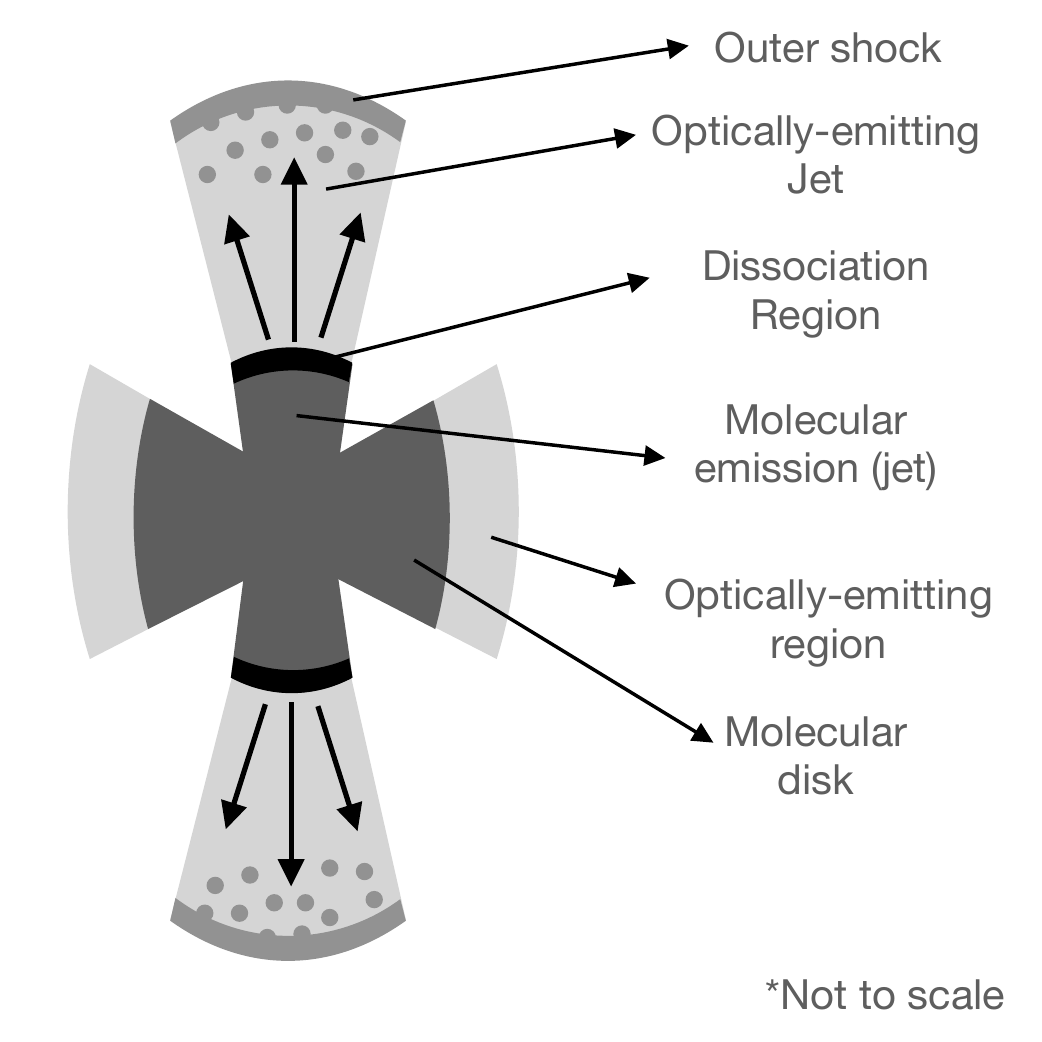}
\caption{
Sketch illustrating the molecular and ionized components of the H-deficient ejecta of A\,58.  
The propagation of the molecular jet component produces a shock that dissociates the molecules (dissociation region), whereas the propagation of the ionized jet component shocks the old H-rich nebula (outer shock). 
}
\label{fig:cartoon}
\end{figure}

\subsection{Stellar Evolution}

To assess the effects of the possible evolution of the ionizing flux from the CSPN of A\,58, the photoionization code Cloudy \citep{Ferland2017} was used to compute a number of synthetic spectra that were accordingly compared to those observed at different epochs. 
The emitting region was assumed to be a shell with density of 2,000 cm$^{-3}$ and outer radius of 0.02 pc (1 arcsec at our distance estimate of 4.3 kpc, see Sec.~\ref{sec:outer_neb}). 
The gas abundances were then adopted to be those reported by \citet{Wesson2008}, whereas the stellar parameters of V605~Aql at a time close to those of the WHT observations were adopted to be $T_\mathrm{eff}$=95,000~K and $L=600$~L$_{\odot}$, similar to those reported in \citet{Clayton2006}, i.e., scaling the luminosity to 4.3 kpc. 
The NLTE PoWR code\footnote{\url{https://www.astro.physik.uni-potsdam.de/~wrh/PoWR/powrgrid1.php}} \citep[see][and references therein]{Hamann2004,Todt2015} was then used to produce an input ionizing flux. 
The inner radius of the emitting region was then varied until the H$\alpha$ and  H$\beta$ fluxes, and the [N\,{\sc ii}] $\lambda\lambda$6548,6584 and [O~{\sc iii}] $\lambda\lambda$4959,5007 line intensity ratios were reproduced.  
This model, however, could not reproduce the intensities of the auroral [N\,{\sc ii}] $\lambda$5755 and [O~{\sc iii}] $\lambda$4363 emission lines, neither those of the [S\,{\sc ii}] $\lambda\lambda$6717,6730, [O~{\sc i}], and [O\,{\sc ii}] emission lines. 
A large number of Cloudy models were then attempted varying both the inner radius of the shell and the effective temperature of a black-body model between 70,000 and 115,000~K to probe both the evolution of the size of the ionized region and effective stellar temperature.  
Whereas these models were able to reproduce the intensity enhancement of the [O~{\sc iii}] and [N~{\sc ii}] emission lines, they under-predicted largely the intensity of the [S~{\sc ii}] lines and failed notably to reproduce the time evolution of the [O~{\sc ii}] doublet.  
In this simplistic model for the physical structure of the ejecta and input ionizing flux, the time evolution of the size of the ionized region and effective stellar temperature can not fully explain the observed line intensity ratio variations.  

\subsection{Shocks}

It shall be noticed that the most problematic emission lines for the photoionization models, i.e., the [O~{\sc ii}] and [S~{\sc ii}] emission lines, are very prone to the effects of shocks.  
As a matter of fact, the [O\,{\sc iii}]/H$\beta$, [N\,{\sc ii}]/H$\alpha$, and [S\,{\sc ii}]/H$\alpha$ line ratios observed at different epochs in the central knot of A\,58 occupy loci associated with shocks in BPT ionization diagrams \citep[][]{BPT1981}.  
This is also the case if these ratios were to be corrected by the low H content implied by the abundances measured by \citet{Wesson2008}.
These authors actually discarded the relevance of shocks arguing that the mechanical luminosity of the stellar wind was a tiny fraction of the stellar luminosity.  
This is certainly the case, as it is also that the fast stellar wind ($v_\infty = 2,500$ km~s$^{-1}$) is not responsible of the shocks resulting in the line ratios observed in the central knot of A\,58.  
These are otherwise indicative of much smaller velocity shocks that would be associated with the 180--270 km~s$^{-1}$ expansion of the H-poor ejecta \citep{Pollacco1992,Tafoya2022}.

We therefore propose that shocks contribute significantly to the total emission observed in the central knot of A\,58, being responsible for the time evolution of its emissivity and ionizing stage.  
The recent detection of a fast ($\sim$280~km~s$^{-1}$) bipolar molecular outflow at the heart of A\,58 \citep{Tafoya2022} with an apparent kinematic age ($\lesssim$20~yr) much smaller than that of the VLTP event ($\approx$100~yr) suggests the concurrent presence of molecular and ionized components in this outflow.  
This is illustrated in the scheme of the collimated outflow of A\,58 shown in Figure~\ref{fig:cartoon}, which is reminiscent of the early evolution of a common envelope with a companion just after the main star experienced a VLTP \citep{RG2022} 
inspired by the recent suggestion of a cooler companion of the CSPN of the born-again PN A\,30 \citep{Jacoby2020}. 

The molecular outflow is still been launched and, as it shocks material previously ejected, molecules would be dissociated and incorporated to the ionized component.  
Meanwhile the ionized component itself would be propagating an outer shock through the old nebula.  
As these two shocks proceed, the mass of the ionized gas increases, and thus its emissivity, whereas the emission from species sensitive to the velocity regime of the shock, such as O$^{++}$, O$^+$, and N$^+$, is enhanced.

\section{Final Remarks}

The analysis of multi-epoch spectroscopic observations of the outer shell and central knot of the born-again PN A\,58 spanning over two decades has revealed a clear brightening and spectral evolution of the latter. 
We note that the multi-epoch spectroscopic observations do not probe exactly the same spatial regions, but the apertures used to extract spectra at each epoch have been carefully selected to register regions with  similar properties. 
At any rate, the observed spectral changes are too large to be explained by this alone.

Finally we note that the chemical abundances derived for the central knot of A\,58 would vary notably depending on whether the emission originates from shocked or photoionized plasmas \citep[see, for instance,][]{Borja2022}.  
If shocks were indeed the prevalent excitation mechanism, the current abundances estimates \citep{Guerrero1996,Wesson2008} would need to be revisited accordingly.  
This would require a major model effort (physical structure and velocity field of the ejecta, shock models, ...) that is besides the scope of this work (Rodr\'\i guez-Gonz\'alez et al., in preparation) and 
that should account for the different spatial coverage of the available multi-epoch spectroscopic observations. 
\\
\\

B.M.M. and M.A.G. are funded by the Spanish Ministerio de Ciencia, Innovaci\'on y Universidades (MCIU) grant PGC2018-102184-B-I00, co-funded by FEDER funds. B.M.M. and M.A.G. acknowledge support from the State Agency for Research of the Spanish MCIU through the ‘Center of Excellence Severo Ochoa’ award to the Instituto de Astrof\'\i sica de Andaluc\'\i a (SEV-2017-0709). J.A.T. thanks funding by Fundación Marcos Moshinsky (Mexico) and the DGAPA UNAM project IA101622. J.B.R.G. acknowledges Consejo Nacional de
Ciencia y Tecnología (CONACyT, Mexico) for a student scholarship. The authors thank D. Tafoya for fruitful discussion during the preparation of the manuscript. We also thank H. Todt for providing us with the stellar atmosphere model of V\,605 Aql. This work made use of IRAF, distributed by the National Optical Astronomy Observatory, which is operated by the Association of Universities for Research in Astronomy under cooperative agreement with the National Science Foundation. 
This work has made extensive use of NASA’s Astrophysics Data System. 

\software{IRAF, Cloudy, PoWR}
\facilities{ORM NOT, ORM WHT, CAHA 3.5m, HST}




\end{document}